\documentclass[11pt,superscriptaddress,floatfix]{revtex4-2}
\usepackage{graphics,amssymb,amsmath,epsfig,color}
\usepackage{mathtools} 
\usepackage{graphicx}
\usepackage{setspace}
\usepackage{hyperref}
\usepackage{braket}

\begin{document}
\onehalfspacing

\title{Unraveling vibronic interactions in molecules functionalized with optical cycling centers}

\author{Paweł W{\'o}jcik}
\thanks{These authors contributed equally.}
\affiliation{Department of Chemistry, University of Southern California, Los Angeles, California 90089, USA}
\author{}
\thanks{Present address: Department of Chemistry and Biochemistry, Florida State University, Tallahassee, Florida 32306, USA} 
\noaffiliation

\author{Haowen Zhou}
\thanks{These authors contributed equally.}
\affiliation{Department of Physics and Astronomy, University of California, Los Angeles, California 90095, USA}

\author{Taras Khvorost}
\affiliation{Department of Chemistry and Biochemistry, University of California, Los Angeles, California 90095, USA}

\author{Guo-Zhu Zhu}
\altaffiliation[Present address: ]{Department of Chemistry, Fudan University, Shanghai 200438, China}
\affiliation{Department of Physics and Astronomy, University of California, Los Angeles, California 90095, USA}

\author{Guanming Lao}
\altaffiliation[Present address: ]{Pritzker School of Molecular Engineering, University of Chicago, Chicago, IL 60637, USA}
\affiliation{Department of Physics and Astronomy, University of California, Los Angeles, California 90095, USA}

\author{Justin R. Caram}
\affiliation{Department of Chemistry and Biochemistry, University of California, Los Angeles, California 90095, USA}

\author{Anastassia N. Alexandrova}
\affiliation{Department of Chemistry and Biochemistry, University of California, Los Angeles, California 90095, USA}
\affiliation{Center for Quantum Science and Engineering, University of California, Los Angeles, California 90095, USA}

\author{Eric R. Hudson}
\affiliation{Department of Physics and Astronomy, University of California, Los Angeles, California 90095, USA}
\affiliation{Center for Quantum Science and Engineering, University of California, Los Angeles, California 90095, USA}
\affiliation{Challenge Institute for Quantum Computation, University of California, Los Angeles, California 90095, USA}

\author{Wesley C. Campbell}
\affiliation{Department of Physics and Astronomy, University of California, Los Angeles, California 90095, USA}
\affiliation{Center for Quantum Science and Engineering, University of California, Los Angeles, California 90095, USA}
\affiliation{Challenge Institute for Quantum Computation, University of California, Los Angeles, California 90095, USA}

\author{Anna I. Krylov}
\affiliation{Department of Chemistry, University of Southern California, Los Angeles, California 90089, USA}
\date{\today}
\bigskip

\begin{abstract}
  We report detailed characterization of the vibronic interactions  between the first two electronically excited states, $\tilde{A}$ and $\tilde{B}$, in SrOPh (Ph = phenyl, -$\text{C}_6\text{H}_5$) and its deuterated
  counterpart, SrOPh-d$_5$ (-$\text{C}_6\text{D}_5$). The vibronic interactions, which arise due to non-adiabatic coupling between the two electronic states, mix the $\tilde{B},\nu_0$ state with the energetically close vibronic level $\tilde{A},\nu_{21}\nu_{33}$, resulting in extra transition probability into the latter state.
  This state mixing is more prominent in the deuterated molecule because of the smaller energy gap between the interacting states. We model the mixing of the $\tilde{A}$ and $\tilde{B}$ states using the K\"oppel--Domcke--Cederbaum (KDC) Hamiltonian parametrized in the diabatic framework of   Ichino, Gauss, and Stanton on the basis of  equation-of-motion coupled-cluster calculations. The simulation attributes the observed mixing to a second-order effect mediated by   linear quasi-diabatic couplings between the $\tilde{A}$-$\tilde{C}$ and $\tilde{B}$-$\tilde{C}$ states.
  Based on the measured spectra, we deduce an effective coupling strength of $\sim0.5$~cm$^{-1}$.
  Non-adiabatic couplings between different electronic states is an important factor that should be considered in the design of laser-cooling protocols for complex molecules.
\end{abstract}
\maketitle


\section{Introduction}

Cold and ultracold molecules are essential for studying detailed molecular interactions and
reactions~\cite{carr_cold_2009, bohn_cold_2017, hu_direct_2019, gregory_moleculemolecule_2021, chen_ultracold_2024}, building new platforms for quantum information science~\cite{bohn_cold_2017, yu_scalable_2019, hutzler_polyatomic_2020, holland_-demand_2023, demille_quantum_2024, cornish_quantum_2024}, and advancing precision measurements
in search of new physics beyond the Standard Model~\cite{isaev_laser-cooled_2010, augenbraun_molecular_2020, augenbraun_laser-cooled_2020, anderegg_quantum_2023, mitra_quantum_2022}.
Laser cooling, which relies on the repeated scattering of photons to slow down the species, has been the
vehicle for achieving such cold temperatures.
However, compared to atoms, the many vibrational degrees of freedom of molecules create undesirable branching pathways, preventing the closure of a repeated optical cycle.
Over the last two decades, tremendous efforts have been made to extend this technique to more complex species.
The most successful accomplishments in laser cooling were based on a 
family of molecules---first proposed by Isaev and Berger\cite{isaev_polyatomic_2016} 
---supporting 
atomic-like states realized in alkaline earth metal atoms attached to electronegative scaffold containing oxygen or halogen~\cite{fitch_laser-cooled_2021, aieta_efficient_2016, baum_1d_2020, collopy_3d_2018, mitra_direct_2020, lasner_magneto-optical_2025}. In these systems, one electron from the metal is donated to an ionic bond with the scaffold and the second electron, localized on the metal, gives rise to atomic-like states. Due to their localized character,
the transitions between these states result in minimal changes to the molecular structure.
Such alkaline earth (I)-oxygen groups can be attached to a variety of scaffolds and
are commonly referred to as an optical cycling center (OCC).
Functionalization with OCC is an effective way for developing versatile chemical platforms with the potential capability for laser cooling ~\cite{kozyryev_proposal_2016, klos_prospects_2020, dickerson_franck-condon_2021, zhu_functionalizing_2022, augenbraun_direct_2023, lao_bottom-up_2024, zhou2025vibronic}.

\begin{figure}[h!]
  \begin{center}
    \includegraphics[width=8 cm]{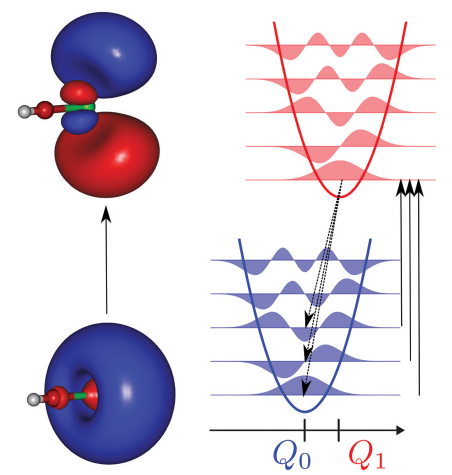}
  \end{center}
    \caption{
      Low-lying electronic states in OCC-functionalized molecules illustrated by the Dyson
      orbitals in SrOH and a cartoon illustrating vibrational wavefunction overlaps between two electronic states.
      Reproduced from Ref. \citenum{Ivanov:MFOCC:19} with permission from the Royal Society of Chemistry.
      \label{fig:1a}}
\end{figure}

The most important criterion for successful optical cycling in molecules is the low branching ratios into
excited vibrational modes. 
In the OCC-functionalized molecules, this criterion is satisfied owing 
to the similar structures of the ground and excited electronic states, which results in nearly
``diagonal'' Franck--Condon factors (FCFs, the overlaps between the vibrational levels)---that is,
only the transitions that do not change the vibrational quantum number are allowed\cite{isaev_polyatomic_2016,Ivanov:MFOCC:19}, as illustrated in Fig.~\ref{fig:1a}.

Within the Born--Oppenheimer framework, calculations of FCFs quantify the  vibrational branchings,
providing a useful guide in the search of suitable molecular
candidates for optical cycling\cite{Herzberg:Diatomic,isaev_polyatomic_2016,Tarbutt:laser:18, Ivanov:MFOCC:19,klos_prospects_2020,Maxim:PCCPCations:2020,Cheng:accurateFCFs:20,Maxim:CoolingLarge:20,Maxim:CoolingLarge:20err,dickerson_franck-condon_2021,Dickerson:OCCarenes:2021,Wojcik:Cations:2022,Mitra:OCCarenes:2022, zhu_functionalizing_2022,Dickerson:saturatedOCC:2022,Augenbraun:directPolyatomic:2023,Khvorost:dualOCC:2024,Pawel:AB:2025}.
However, higher-order terms and additional interactions beyond the Born--Oppenheimer approximation can change the predicted vibrational branching ratios.
This can be especially problematic for larger molecules with many vibrational modes.

\begin{figure}[h!]
    \begin{center}
        \includegraphics[width=7cm]{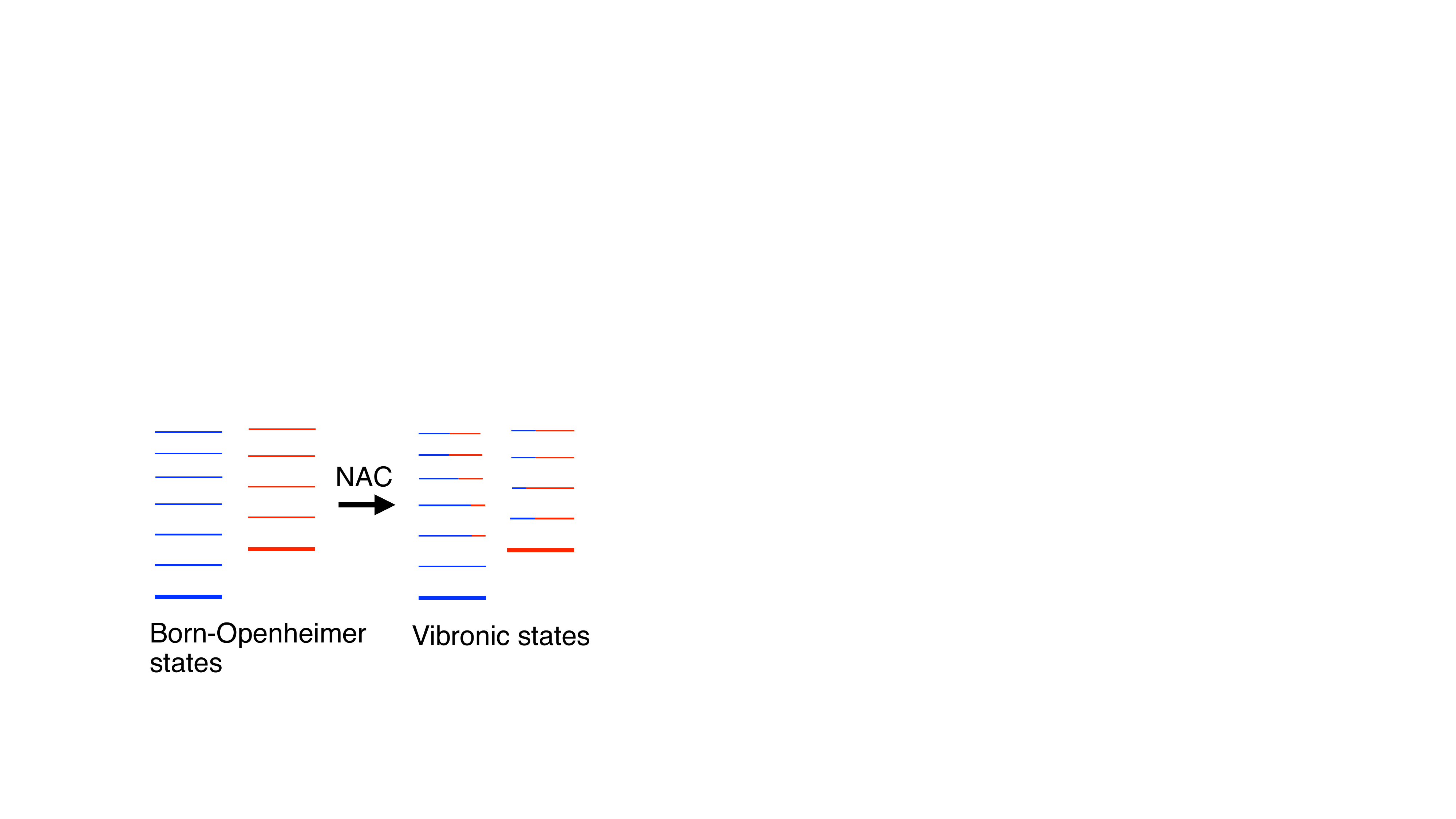}
    \end{center}
    \caption{
 Non-adiabatic coupling mixes vibrational levels from different electronic states, which are
 non-interacting within Born--Oppenheimer approximation.
 The mixed vibronic states lead to more decay pathways.
    \label{fig:1b}}
\end{figure}

Vibronic effects originate in the non-adiabatic couplings (NACs) between
the Born--Oppenheimer electronic states, as illustrated in Fig. \ref{fig:1b}.
In the Born--Oppenheimer approximation, each electronic state supports its manifold of vibrational states, and the vibrational states from the different electronic states do not interact. The NACs couple these zero-order states, resulting in the manifold of vibronic states that have contributions from different
electronic states.

\begin{figure}[h!]
    \begin{center}
        \includegraphics[width=14cm]{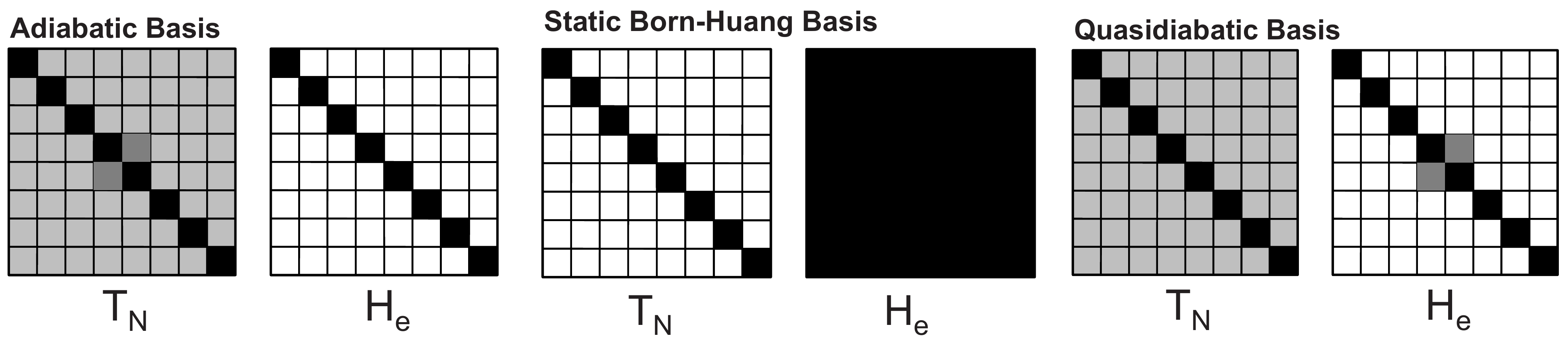}
    \end{center}
    \caption{Schematic illustration of molecular Hamiltonian in the adiabatic (or Born--Oppenheimer)
      basis, naive diabatic (here, static Born--Huang) basis, and quasi-diabatic basis.
   $T_N$ represents the nuclear kinetic energy operator and $H_e$ is the electronic Hamiltonian.
      In the quasi-diabatic basis, the states are close to the adiabatic states and
      the small derivative couplings remain, but the dominant couplings are now appear as off-diagonal terms
      in the electronic Hamiltonian.
Reproduced from Ref. \citenum{Stanton:EOMIPdeg:09}, with the permission of AIP Publishing.
    \label{fig:Diab}}
\end{figure}

NACs, often referred to as derivative couplings, arise because the
electronic
wavefunctions change their character upon nuclear motions---they are
zero  if the electronic wavefunctions do not change upon nuclear displacements.
The electronic states that do not change their character upon nuclear
displacements are called diabatic states. In the basis of diabatic states,
the derivative couplings are zero, but these states---which are no longer are
eigenstates of the electronic Hamiltonian---are coupled by the off-diagonal
elements called diabatic couplings. Although the diabatic framework is not
uniquely defined, it has advantages in the context of
computational modeling of vibronic effects---it
avoids the problem of diverging couplings (common in the calculations of NACs)
and facilitates effective parameterization of vibronic
Hamiltonians.
Here we use the quasi-diabatic
framework of Ichino, Gauss and Stanton\cite{Stanton:EOMIPdeg:09},
which is particularly well-suited for setting up vibronic Hamiltonians. In this
framework, illustrated in Fig. \ref{fig:Diab}, the diabatic states correspond to the adiabatic states 
computed at a reference geometry and coupled by non-adiabatic coupling
force (the main ingredient of the NAC) computed at this geometry; these couplings are equated to diabatic couplings. 
Since the derivative couplings are not fully removed, the framework is called
quasi-diabatic\cite{Stanton:EOMIPdeg:09,Shirin:NAC:17}. 
This framework was designed to enable parameterization of the 
K\"oppel-Domcke-Cederbaum (KDC) Hamiltonian---a molecular Hamiltonian expressed in the basis of diabatic states coupled by off-diagonal matrix elements.\cite{Cederbaum:LVC:84, KDC:81, Koppel:CIbookCh7:04} The diagonalization of the KDC Hamiltonian yields coupled vibronic states, such as those shown in Fig. \ref{fig:1b}, facilitating the calculations of the transitions between these levels. The details of the KDC Hamiltonians and the way of their
parameterization are described below.

In a recent work~\cite{forreviewpaper}, we have shown
the higher electronic excited states of CaOPh and SrOPh (Ph = $\text{-C}_6\text{H}_5$, phenyl)
are strongly coupled to the vibrational levels of the lower electronic states.
The vibronic coupling mixes vibrational levels from the two electronic manifolds, thereby introducing
additional decay pathways from the mixed components (see Fig.~\ref{fig:1b}). These vibronic interactions render  optical cycling from  higher-lying electronic excited states
ineffective. The experimental observations of vibronic effects have been supported by the
simulations. The theory was in qualitative agreement with the experiment, however,
a state-specific comparison between theory and experiment was not possible due to the large number of coupled vibronic states at high energies.
In the present work, we focus on the lowest two electronically excited states, $\tilde{A}$ and $\tilde{B}$,
of SrOPh.
Because of the small energy separation ($\sim300 \text{ cm}^{-1}$), only a few vibrational levels need to be included in the model, making this system an excellent testbed for  comparing theory with experiment.  
We were able to characterize vibronic couplings
between the $\ket{\tilde{B},\nu_0}$ state and the nearby vibrationally excited level of the lower electronic state $\ket{\tilde{A},\nu_{21}\nu_{33}}$.
The nature of the coupling is further confirmed by using the deuterated compound SrOPh-d$_5$ with a much closer energy separation between the two states, which results in a stronger state mixing.
Based on the measured spectra, we obtain an effective coupling strength of $\sim0.5 \text{cm}^{-1}$ between these two states, consistent for both undeuterated and deuterated species.
The observed spectra can be only reproduced by calculations using the KDC vibronic Hamiltonian.
The theory reveals that the coupling between the $\tilde{A}$ and $\tilde{B}$ states is a second-order effect that involves the linear coupling with the higher $\tilde{C}$ state.
This effect is analogous to the spin--orbit vibronic coupling  mechanism in the linear CaOH molecule between the $A_{1/2}$ and $A_{3/2}$ states~\cite{zhang_accurate_2021, zhang_intensity-borrowing_2023}.
Our study reveals the details of vibronic interactions in SrOPh and illustrates the necessity to consider NACs when discussing prospects of laser cooling of large molecules.

The structure of the paper is as follows: Sec.~\ref{sec:methods} describes the theoretical and experimental methods, Sec.~\ref{sec:resuts} presents the results and discussion, and Sec.~\ref{sec:conculsions} gives the concluding remarks.

\section{Methods}
\label{sec:methods}
\subsection{Theoretical Methods}

Similar to other molecules from this family, SrOPh~\cite{lao_laser_2022} features 
an alkali-atom-like spectrum due to an ionic Sr-O bond and
a single unpaired electron localized at the metal atom (Fig.~\ref{fig:SrCouplings}a).
The molecular ground state, $\tilde{X}^2$A$_1$, corresponds to the hydrogen-atom-like $^2$S(5$s^1$) state of Rb. 
Molecular states $\tilde{A}^2$B$_{2}$ and $\tilde{B}^2$B$_{1}$ correspond to the two components of the
triply degenerate $^2$P(5$p^1$) atomic states, but the degeneracy is lifted due to lower-symmetry molecular field.
The $\tilde{C}^2$A$_1$ state corresponds to the third component of $^2$P state,
with the Dyson orbital  pointing towards the
phenoxide moiety; this state shows noticeable hybridization and
has higher energy. 
All four states have nearly identical equilibrium structures and
nearly parallel  potential energy surfaces.
This similarity in the electronic states results in nearly diagonal FCFs, which, in turn, result in he vibrational-state-preserving radiative decays, making
these OCC-functionalized species useful for laser cooling (see Fig.~\ref{fig:SrCouplings}a).

We carried out vibronic simulations using the KDC Hamiltonian.~\cite{Cederbaum:LVC:84, KDC:81, Koppel:CIbookCh7:04}
The KDC Hamiltonian is represented in a basis of diabatic states, coupled by the
off-diagonal elements (diabatic couplings). We parametrize the KDC Hamiltonian 
using the quasi-diabatic framework  by
Ichino, Gauss, and Stanton.~\cite{Stanton:EOMIPdeg:09} 

The KDC Hamiltonian is built in the basis of 3 electronic states and is expanded in the dimensionless normal coordinates of the ground state:
\begin{equation}
    H _{KDC} = 
    H _0 \mathbf{1}
    +
    \begin{bmatrix}
    V ^{(\tilde{A})} & V ^{(\tilde{A}\tilde{B})} & V ^{(\tilde{A}\tilde{C})} \\
    V ^{(\tilde{B}\tilde{A})} & V ^{(\tilde{B})} & V ^{(\tilde{B}\tilde{C})} \\
    V ^{(\tilde{C}\tilde{A})} & V ^{(\tilde{C}\tilde{B})} & V ^{(\tilde{C})} \\
    \end{bmatrix}.
    \label{eq:kdc}
\end{equation}
It consists of the diagonal harmonic terms
\begin{equation}
    H _0 = -\frac{1}{2} \sum _{i = 1} ^ {N _{nm}} 
    \omega _i 
    \left(
    \partial ^2 _{Q _i}
    + 
    Q _i ^ 2
    \right),
    \label{eq:kdc_H0}
\end{equation}
the diagonal diabatic potential terms
\begin{equation}
    V ^{(\alpha)} 
    = 
    E ^{(\alpha)}
    +
    \sum _{i = 1} ^ {N _{nm}} 
    \kappa _i ^{(\alpha)}
    Q _i,
    \label{eq:kdc_Vdiag}
\end{equation}
and the off-diagonal diabatic coupling terms
\begin{equation}
    V ^{(\alpha\beta)} 
    = 
    \smashoperator{\sum _{i \in \{\text{coupling modes}\}}}
    \lambda _i ^{(\alpha \beta)}
    Q _i.
    \label{eq:kdc_Voff}
\end{equation}
In the above, $N_{nm}$ is the number of normal modes included in the model; $\mathbf{1}$ is a $3\times 3$ identity matrix; $\omega_i$ is the ground-state harmonic frequency of the normal mode $i$; $Q_i$ is a dimensionless normal coordinate; $E^{(\alpha)}$ is the vertical excitation energy of the electronic state $\alpha$; $\kappa ^{(\alpha)} _i$ is energy gradient of state $\alpha$ along the dimensionless normal mode $i$; $\lambda ^{(\alpha \beta)} _i$ is the linear diabatic coupling constant between states $\alpha$ and $\beta$ along normal mode $i$.
In the quasi-diabatic
framework of Ichino,  Gauss, and Stanton\cite{Stanton:EOMIPdeg:09},
the linear diabatic coupling is computed as the non-adiabatic force at
the reference geometry (here, the ground-state equilibrium geometry).
This coupling determines the strength of the vibronic interactions. 

We parametrized the KDC Hamiltonian using the coupled-cluster (CC)
and equation-of-motion CC (EOM-CC) methods,~\cite{Bartlett:CC_review:07,Krylov:EOMRev:07} following
the same protocols as in Refs. \citenum{Wojcik:ozone:2024,Pawel:Pyrazine:2025}.
Specifically, we used EOM-CC for electron attachment starting from closed-shell cationic reference state. To visualize electronic states, we used Dyson orbitals\cite{Oana:Dyson:07,Krylov:Orbitals} computed 
using many-body EOM-CC wavefunctions.
We computed vertical excitation energies $E ^{(\alpha)}$ using composite schemes involving basis-set extrapolation and EOM-CC with single and double
excitations (EOM-CCSD) as well as EOM-CCSD with perturbative triples correction
(EOM-CCSD$^*$).~\cite{StantonGauss:SD3:96} 
We note that the highest level of theory has most significant effect
on the $\tilde{C}$ state. To evaluate the effect of spin--orbit interaction, we 
used state-interaction approach with spin--orbit couplings computed as the matrix elements of the Breit--Pauli Hamiltonian using EOM-CCSD wavefunctions\cite{Pokhilko:SOC:19}, similar to our previous study\cite{Khvorost:dualOCC:2024}.
The parameters of the KDC Hamiltonian, Eq.~\eqref{eq:kdc}, were expanded around the optimized geometry of the $\tilde{X}$ state.
The linear diabatic coupling constants, $\lambda ^{(\alpha \beta)}_i$,
were computed as NAC forces
between the respective EOM-CCSD electronic states at
the reference geometry.~\cite{Stanton:EOMIPdeg:09,Shirin:NAC:17}
All CC and EOM-CC calculations were carried out using \textsc{CFOUR} and \textsc{Q-Chem}.~\cite{cfour, cfour:2020, qchem_feature, qchem5_full}

We then simulated the vibronic spectrum with the \textsc{xsim} program.~\cite{Sharma:xsim_socjt:2024}
The vibronic simulations used 15 basis functions per vibrational mode and 2000 Lanczos iterations. 
The methods used in this work were recently successfully applied to
simulate vibronic effects in other molecules, e.g., YbOH, CaOH, SrOH, RaOH, SrOCH$_3$, SrNH$_2$, NO$_3$, benzene cation, O$_3$, and pyrazine.~\cite{Doyle:YbOH:20, zhangAccuratePredictionMeasurement2021,
Doyle:SrOH:22, zhangIntensityborrowingMechanismsPertinent2023,
frenettVibrationalBranchingFractions2024, Stanton:NO3:07, Koppel:02,
Wojcik:ozone:2024,Pawel:Pyrazine:2025}

In our model,
the multi-state multi-mode KDC Hamiltonian included three excited electronic states:
$\tilde{A}$, $\tilde{B}$, and $\tilde{C}$. Because including 
all 33 normal modes is computationally intractable, we considered 
a varying number of normal modes, in order to assess the impact of various modes on the vibronic spectrum. 

The molecular orientation and normal mode labels follow the Mulliken's convention.~\cite{Mulliken:55:symnot} 
The molecule is placed in the $yz$-plane with the symmetry axis along the $z$ axis.
The normal modes are ordered first by their symmetry: $a_1$, $a_2$, $b_1$, $b_2$. 
Within each symmetry block the modes are ordered by their harmonic frequencies in a decreasing order. 
Out of the 33 normal modes of the molecule, 12 transforms with $a _1$ irrep, 3 with $a_2$, 7 with $b _1$, and 11 with $b _2$, see Fig.~\ref{fig:SrCouplings}.

\subsection{Experimental Methods}

The experimental setup have been described in our previous publications~\cite{zhu_extending_2024}. In brief, the vibrational branching pathways were determined by dispersed laser-induced fluorescence (DLIF) spectroscopy inside a cryogenic buffer gas cell. SrOPh molecules were produced by reacting Sr metal with phenol vapor in the cryogenic cell held at $\sim$20K cooled by Ne buffer gas. The phenol vapor was introduced into the cryogenic cell through a heated gas line originating from a heated reservoir. The metal atoms were introduced by laser ablating using an Nd:YAG laser (Minilite) with a pulse energy of approximately 6 mJ. All chemicals were purchased from Sigma-Aldrich without further purification. The produced SrOPh molecules were then excited by a tunable pulsed dye laser (LiopStar-E, linewidth 0.04~cm$^{-1}$). The resulting fluorescence was collected into a grating monochromator (McPherson 2035) and detected using an ICCD camera (Andor iStar 320T). 

In order to verify vibronic interactions between different electronic states, we  performed the DLIF spectroscopy at varying excitation wavelengths while simultaneously monitoring the fluorescence detection, resulting in 2D spectra (see Fig. \ref{fig:LIF} for example).
The 2D spectra can be analyzed along the two dimensions, producing either DLIF spectra at fixed excitation wavelengths or excitation spectra at fixed fluorescence emission wavelengths.
In the excitation spectrum, the ground-state molecules were excited into various vibrational levels of a higher electronic state.
Because of similar electronic structures of the OCC-functionalized molecule, the decay from these vibrationally excited states into the ground state typically preserves their vibrational levels. 
As such, the excitation spectrum resembles the DLIF spectrum within the Born--Oppenheimer approximation.
However, in the presence of vibronic interactions, additional transition probabilities arise from the coupled components of a different electronic state.
By comparing the excitation and DLIF spectra in a 2D spectroscopy measurement, we were able to identify additional peaks of vibronic nature.


\section{Results and discussions}
\label{sec:resuts}

\subsection{Theoretical results}
    
\begin{figure}[h!]
    \begin{center}
        \includegraphics[width=12 cm]{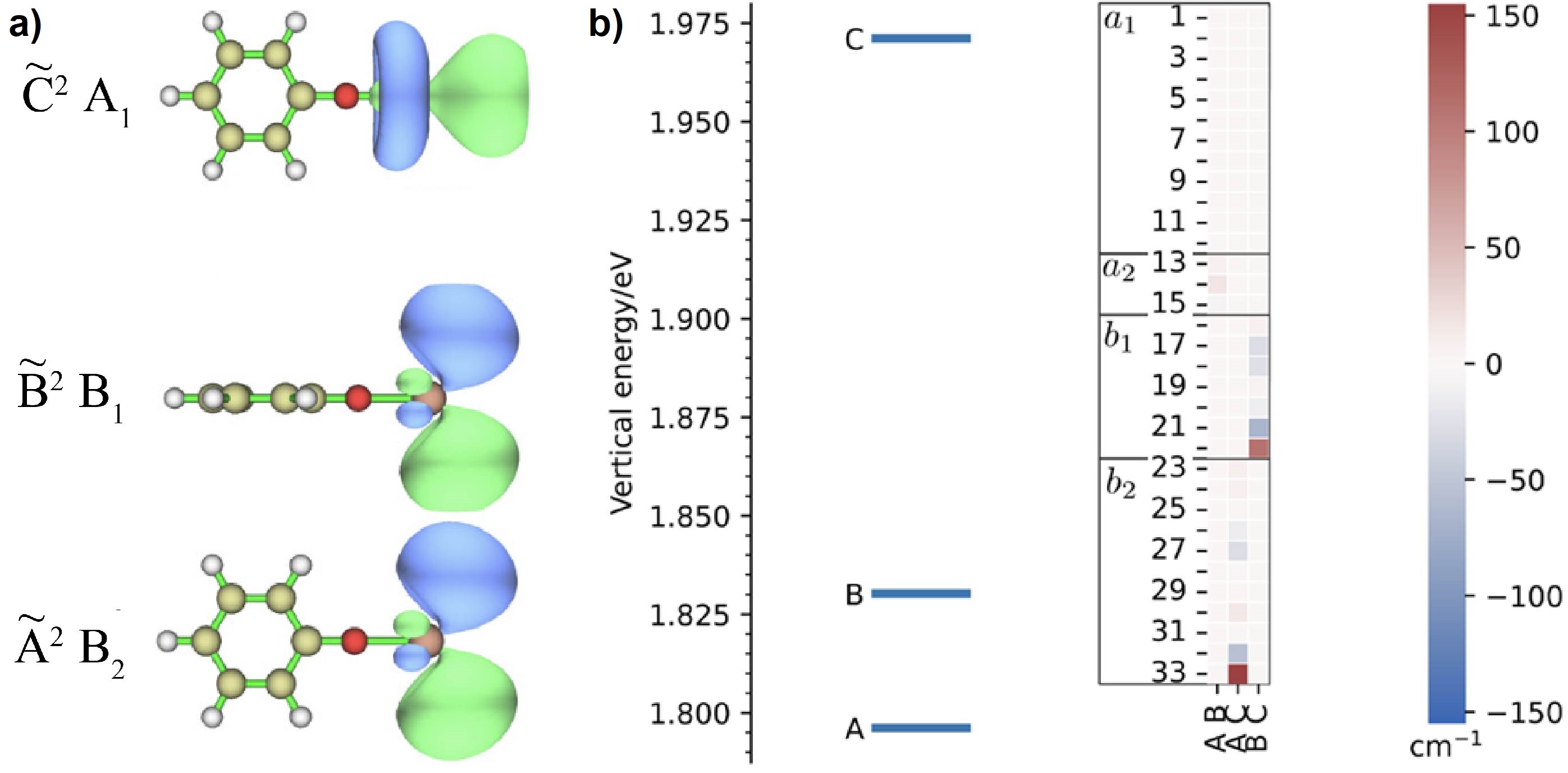}
    \end{center}
    \caption{SrOPh.  
        (a) Dyson orbitals. (b) Vertical excitation energies ($E ^{(\alpha)}$ from Eq.~\eqref{eq:kdc_Vdiag}) and the linear diabatic couplings ($\lambda$ from Eq.~\eqref{eq:kdc_Voff}). 
    \label{fig:SrCouplings}}
\end{figure}

We begin with a discussion of the KDC Hamiltonian, focusing on the most
important parameters of the model. 
The harmonic frequency of the Sr-O stretch, $\nu _{12} (a_1)$, is 239~cm$^{-1}$. The harmonic frequencies of the non-fully-symmetric modes are: 65~cm$^{-1}$ for the in-plane Sr-O-C bend, $\nu _{33}(b_2)$ and 84~cm$^{-1}$ for the out-of-plane Sr-O-C bend, $\nu_{22} (b _1)$. The next mode, in the order of increasing frequency, $\nu _{21} (b _1)$, has the frequency 273~cm$^{-1}$ and can also be described as an out-of-plane Sr-O-C bend. 
These four modes, their combinations, and their couplings produce most of the vibronic features in the region between and close to the
$\tilde{A}$, and $\tilde{B}$ states, which  are separated
vertically by only $275$~cm$^{-1}$ in our model.

Figure~\ref{fig:SrCouplings} shows the relevant electronic states and
the structure of the linear quasi-diabatic couplings in SrOPh. By symmetry,  $\tilde{A}$ and $\tilde{B}$
can be coupled by $a_2$ modes, $\tilde{A}$ and $\tilde{C}$ by $b_2$ modes, and
 $\tilde{B}$ and $\tilde{C}$ by $b_1$ modes.
The couplings between the $\tilde{A}$ and $\tilde{C}$ states (middle column) are the strongest along the in-plane Sr-O-C bending modes; these modes move
Sr and O towards the region where electronic density of the unpaired electron in the $\tilde{A}$ state is. Similarly, the $\tilde{B}$ and $\tilde{C}$ states couple along the out-of-plane Sr-O-C modes. The linear vibronic couplings
between $\tilde{A}$ and $\tilde{B}$ states are vanishingly small. The coupling between these two closely
lying states appear in this model in the second order, i.e., through  vibronically active combination modes. A complete list of the model parameters is given in  Sec.~\ref{si:sec:comp_details} in the SI.

\begin{figure}[h!]
    \begin{center}
        \includegraphics[width=\linewidth]{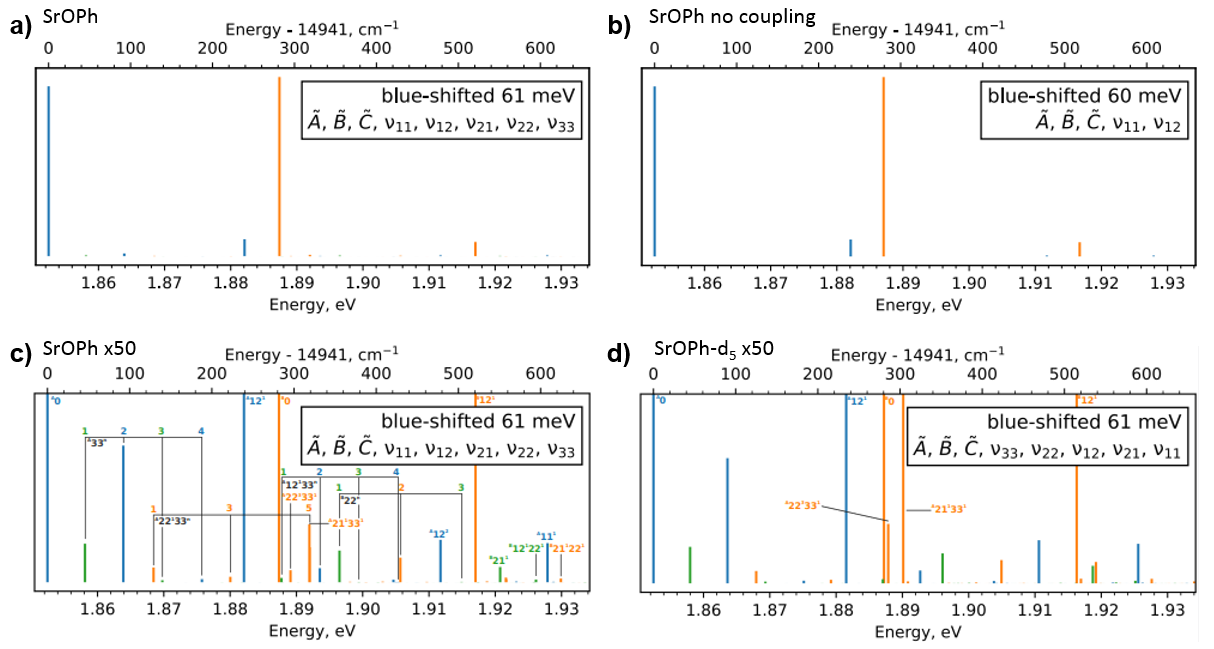}
    \end{center}
    \caption{
      Simulated absorption spectra of SrOPh and SrOPh-d$_5$.
      Colors denote the symmetries of the vibronic peaks: blue B$_2$, orange B$_1$, green A$_1$. The insets
      list the states and modes active in the simulation. The peaks are labeled as $^{S}\nu_{i}^{f}$, where $S$ denotes
      the electronically excited state, $v$ denotes the vibrational mode, $i$ and $f$ define the vibrational quantum number in the excited and ground electronic state. (a) Simulated spectra of SrOPh with vibronic coupling. (b) Simulated spectra of SrOPh without coupling (i.e., Franck--Condon simulation). (c) Zoomed-in version of (a) showing details of peaks resulted from the vibronic couplings. The corresponding vibrational modes are also assigned. (d) Zoomed-in version of SrOPh-d$_5$ spectra simulated with the same model. Deuteration shifts the frequency of the vibrational modes. The increased mixing is a result of smaller energy gap between the vibronically coupled states. 
    \label{fig:SrOPh_theo}}
\end{figure}

To better understand the role of the vibronic effects in the simulated spectrum,
we compare the vibronic spectrum with the simulation that does not include  vibronic couplings.
Figure ~\ref{fig:SrOPh_theo}(a,b) shows the simulated absorption spectra of SrOPh with and without the vibronic coupling. 
The uncoupled spectrum is typical for a laser-coolable molecule. The most prominent peak in the spectrum corresponds to the electronic transition that does not change the vibrational state.
The key transitions that change the vibrational state excite the $\nu_{12}$ mode (Sr-O stretching). The progression in $\nu_{12}$ is present both in the $\tilde{A}$ and $\tilde{B}$ states, but the peaks' intensities make less than
5\% of the 0-0 peak's intensity.
Both simulations also show the small $\nu _{11} (a_1)$ peak at about 600~cm$^{-1}$, which corresponds to the second lowest-frequency mode with the Sr-O stretching character.
The coupled spectrum, however, reveals a massive enhancement in the number of vibronic states that are weakly allowed as a consequence of the vibronic effects.
Figure ~\ref{fig:SrOPh_theo}(c) zooms in to show the finer features of the vibronic spectrum, which are missing in the uncoupled simulation.
The KDC-Hamiltonian-based simulations reveal that there are two mechanisms leading to the appearance of these features: the direct and second-order vibronic couplings.

\subsubsection{The direct coupling of the $\tilde{A}$ and $\tilde{B}$ states to the $\tilde{C}$ state}

The main signature of the vibronic interactions is the appearance of progressions 
resulting directly from the linear vibronic coupling model.
Figure~\ref{fig:SrOPh_theo}c provides a useful guide for the analysis by 
presenting the assignment of the vibronic features of the fully \emph{ab initio} simulated spectrum. 
The assignment uses the designations of the Franck--Condon-type decoupled state, which  mark the leading terms in the otherwise mixed state.
The linear vibronic couplings model used in this work is based on diabatic couplings between the electronic states.
These couplings appear between pairs of electronic states as expansion coefficients along non-fully-symmetric
modes, as shown in Figure~\ref{fig:SrCouplings}. 
Such modes are inactive in the uncoupled Franck--Condon simulation. The couplings enable the mixing of electronic states along the coupling modes. 

The lowest-energy progression of this type appears for the $\nu _{33}$ mode (in-plane Sr-O-C bending) and is labeled $^{\tilde{A}}33^n$ in Figure~\ref{fig:SrOPh_theo}c. 
The symmetry of vibronic peaks in this progression changes between the symmetry of the host electronic state ($\tilde{A}$, for even number of quanta) and the coupled electronic state ($\tilde{C}$, for odd number of quanta). This progression differs from a typical Franck--Condon progression in the peaks' intensity distribution. The intensities depend not only on the number of vibrational quanta but also on the energy separation from the state that leaks the intensity. Analogous progressions are also marked as $^{\tilde{A}} 12 ^1 33 ^n$ and $^{\tilde{B}}22 ^n$
in Figure~\ref{fig:SrOPh_theo}c. The latter corresponds to the coupling of the $\tilde{B}$ and $\tilde{C}$ states along the $\nu _{22}$ mode. Peaks $^{\tilde{B}}12^1 22^1$ and $^{\tilde{B}}21 ^1$ also appear through this mechanism, however, only the first peak of both series is visible in Figure~\ref{fig:SrOPh_theo}(c).

\subsubsection{Second-order coupling between $\tilde{A}$ and $\tilde{B}$ states}

The progressions discussed above arise as a result of the linear coupling present in the KDC Hamiltonian --- new vibronic states appear due to couplings along non-fully-symmetric modes.
Figure~\ref{fig:SrCouplings} shows that the couplings between the $\tilde{A}$ and $\tilde{B}$ states are very small compared to the coupling to the $\tilde{C}$ state.
Despite the absence of a direct, or first-order, coupling, the $\tilde{A}$ and $\tilde{B}$ states couple via a
second-order mechanism. 

Symmetry of the $\tilde{A}$ and $\tilde{B}$ states dictates that these two can only mix along modes of the $a_2$ symmetry. Figure~\ref{fig:SrCouplings} shows that for the lowest-frequency mode of this symmetry [mode $\nu_{15} (a_2)$ with harmonic frequency $426$~cm$^{-1}$] the coupling is vanishingly small,
making it unlikely to produce significant features. There are, however, other modes that have the desired $a_2$ symmetry---these are the combination bands combining vibrations of $b_1$ and $b_2$ symmetry.
The  $b_1$ and $b_2$ modes couple the $\tilde{B}$-$\tilde{C}$ and $\tilde{A}$-$\tilde{C}$ states; 
they exhibit significant vibronic activity, as discussed above. 
The vibronic features appearing along such combination modes can be described as the second-order effects.

The inspection of the shapes of the Dyson orbitals and the normal modes helps to rationalize the magnitudes of the couplings. 
The strongest coupling between the $\tilde{A}$ and $\tilde{C}$ states is by  mode $\nu_{33}$. This mode moves the Sr-O-C moiety in the same plane where the electronic density of the unpaired electron of the $\tilde{A}$ state is --- hence, this motion brings the 
the electronic densities of the unpaired electron in the two states towards each other. 
This observation provides a visual explanation of a strong interaction between the electronic and vibrational degrees of freedom. Similar argument applies to the $\tilde{B}$-$\tilde{C}$ coupling along  mode $\nu_{22}$. 

To continue with this analysis, let us look at the modes of a symmetry that can couple the $\tilde{A}$ and $\tilde{B}$ states, namely, modes $\nu _{13}$, $\nu_{14}$, and $\nu _{15}$, all of the $a_2$ symmetry. These modes do not displace the Sr-O-C moiety---hence, activation of either of the $\nu _{13}$, $\nu_{14}$, or $\nu _{15}$ modes does not affect the part of the molecule where the electronic densities of the two states differ and, consequently, no significant vibronic interaction is expected. However, if one combines a single excitation in one of the planar Sr-O-C vibrations with an out-of-phase contribution, one ends up with a motion where a bent Sr-O-C rotates around its axis---and such motion mixes the two out-of-plane $p$-like densities of the two states. This second-order motion arising from combination bands may be viewed as a qualitative explanation of the effect described below. 

The progression labeled as $^{\tilde{A}}22 ^1 33 ^n$ in
Figure~\ref{fig:SrOPh_theo}c is the second-order effect arising from the $\nu _{33}$ (Sr-O-C in-plane bend) and $\nu_{22}$ (Sr-O-C out-of-plane bend) modes. These vibronic peaks draw the intensity from mixing with the $\tilde{B}$ state. Only odd number of excitations appear, as all of $^{\tilde{B}}22 ^1 33 ^{2n}$ states have the $A_2$ symmetry and are dark in the 0~K absorption spectrum. The peak labeled  $^{\tilde{A}}22^3 33^1$ can also be viewed as the second peak in the unlabeled progression $^{\tilde{A}}22^n 33^1$.

The coupling between the $\nu _{33}$ and $\nu _{22}$ modes is also interesting in comparison with the work on linear triatomics, like CaOH or YbOH.~\cite{li1995high,zhang_intensity-borrowing_2023,jadbabaie2023characterizing} These two modes would be degenerate bends of Sr-O-C, if it was not for the symmetry-breaking phenyl ring. In the case of linear triatomics, the excited states of these bending modes can be expressed in a basis forming vibrational angular momentum. Such basis is often more convenient for discussion of angular momentum couplings.~\cite{pilgramProductionCharacterizationYtterbium2022}
Analogously, the $\tilde{A}$ and $\tilde{B}$ states would correlate with a degenerate pair in the linear molecule limit, however it was already noted that the angular momentum of the $\tilde{A}$ and $\tilde{B}$ states is largely quenched by the symmetry-breaking phenyl ring.~\cite{Augenbraun:CaOPh:2022}
Given this perspective, the coupling of the $\tilde{A}$ and $\tilde{B}$ states through the combined $\nu_{33}$ and $\nu _{22}$ modes can be correlated with the Renner--Teller coupling in linear molecules. 

The second-order effect also manifests in the modulation of the relative intensities of the vibronic features. The peak $^{\tilde{A}} 22^1 33^1$ is more intense than $^{\tilde{A}} 22^1 33^3$, where it seems like the vibrational overlap dominates over the intensity borrowing coming from mixing, while the $^{\tilde{A}} 22^3 33^1$ peak shows higher intensity than $^{\tilde{A}} 22^1 33^3$, likely due to the the very strong $^{\tilde{B}}0$ peak moving closer. 

Another intense peak of the second-order type is $^{\tilde{A}} 21^1 33^1$. The mode $\nu _{21}$ has a significant component of the Sr-O-C out-of-plane bend, similar to $\nu _{22}$. Despite the fact that the coupling along $\nu _{21}$ is weaker than the coupling along $\nu _{22}$, this peak is much more intense than $^{\tilde{A}} 22^1 33^1$, a difference that is explored in the next section.

\subsubsection{Sensitivity of peak features observed by isotope substitution}

The discussion of the second-order vibronic effects highlights that the intensity of the vibronic features is strongly tied not only to the strength of the diabatic couplings but also to relative energy gaps between the coupled states. This effect can be readily observed by comparing  SrOPh with the deuterated molecule SrOPh-d$_5$. SrOPh-d$_5$ is
characterized by the same parameters as SrOPh, except for the changes in the normal modes, effectively allowing us
to study the response of the vibronic spectrum to the shifts in the positions of the uncoupled vibrational states. In the deteurated molecules, the harmonic frequencies of the vibrational modes are lowered by about 5\%. But the characters of the lowest frequency modes remain largely unchanged and can still be identified as various displacements of the Sr-O-C moiety.

Fig. \ref{fig:SrOPh_theo}d shows 
the simulated vibronic spectrum for SrOPh-d$_5$ of the deuterated molecule. The progressions discussed in the previous section remain largely unaffected by the deuteration, except for small frequency shifts of the deuterated modes. However, there are two states that respond to the deuteration much stronger:
the $^{\tilde{A}}22^3 33^1$ and $^{\tilde{A}}21^1 33 ^1$ states, which are the closest to the $^{\tilde{B}}0$ state. For both of them, the small change in the frequency leads to significant increase in their intensity, as the shift moves both states closer to the bright $^{\tilde{B}}0$ peak. This reveals that for some dark
states the vibronic coupling can produce significant intensity borrowing. A similar effect has been observed in our previous works with the Fermi resonance of CaOPh molecule. The   resonance can be diminished by replacing  Ca with Sr, which changes the frequency of the vibration modes.~\cite{zhu_extending_2024}

\subsection{Experimental Results}

\subsubsection{2D DLIF spectroscopy of $\text{SrOPh}$ and $\text{SrOPh-d}_5$}

In the theoretical spectra shown in Fig.\ref{fig:SrOPh_theo}c and \ref{fig:SrOPh_theo}d, many of the peaks produced by vibronic couplings are too weak to be experimentally measured in an excitation spectra.
However, near the strong $^{\tilde{B}}0$ peak, the intensity borrowing from the much favored 0-0 transition leads to significantly increased intensities of the neighboring vibronic peaks. 
In particular, the $^{\tilde{A}}21^1 33 ^1$ peak in the deuterated molecule is the strongest among all additional peaks arising due to non-adiabatic coupling. 
Its comparison with non-deuterated SrOPh further provides a perfect testbed of theoretical models.

\begin{figure}[h!]
    \centering
    \includegraphics[width=\linewidth]{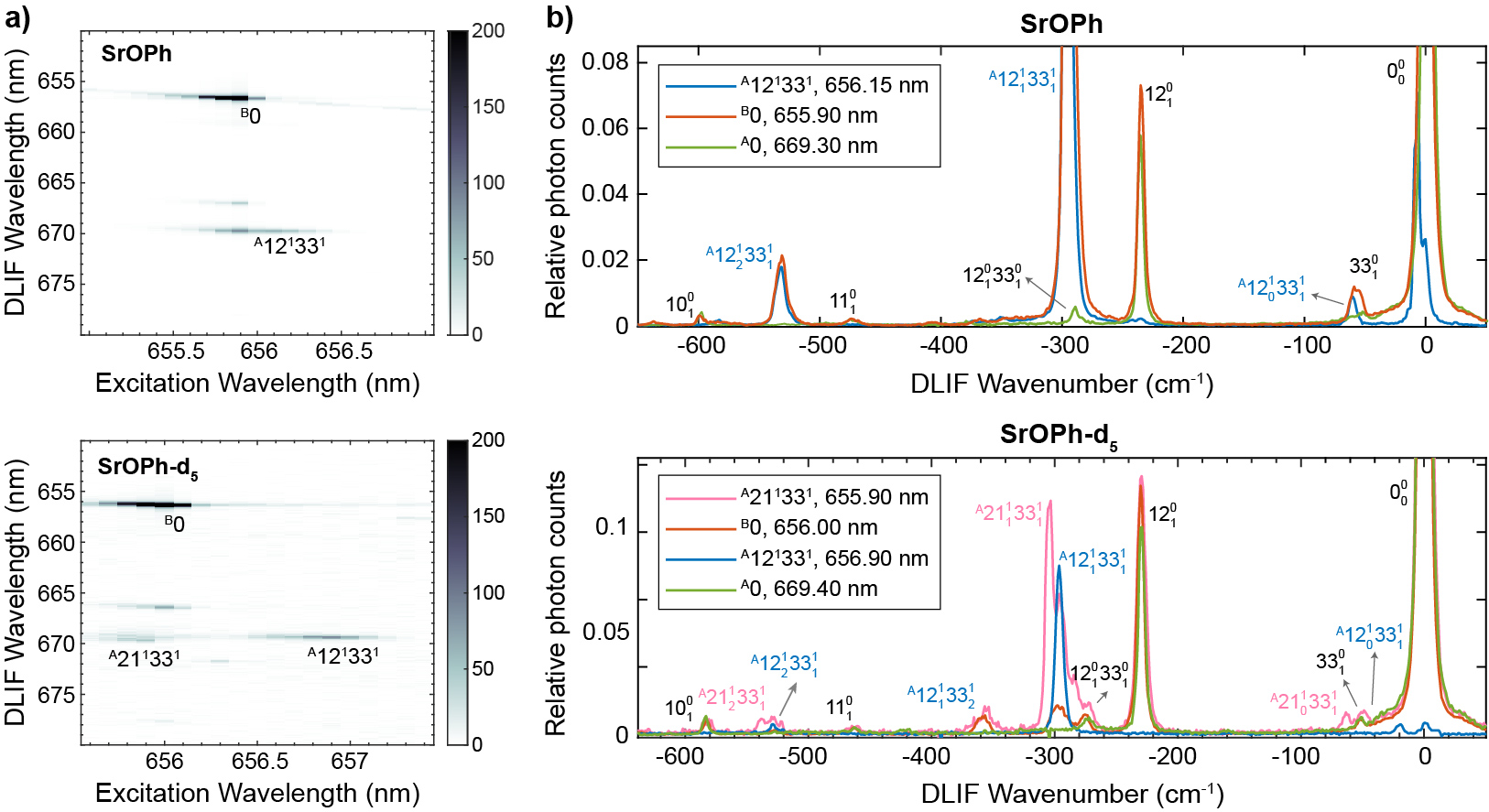}
    \caption{(a) 2D LIF scans for SrOPh and SrOPh-d$_5$. The $\tilde{A}$ and $\tilde{B}$ state excitation peaks can be clearly identified from their vertical relaxation. Notably, an additional excitation peak ($^{\tilde{A}}21^1 33 ^1$) of the $\tilde{A}$ state is observed in SrOPh-d$_5$. We attribute this peak to the vibronic coupling because of closer energy separations with $^{\tilde{B}}0$. (b) Comparison between the DLIF spectra from the $^{\tilde{B}}0$ state and the nearby $\tilde{A}$ state vibrations. The spectra are normalized and overlapped on top of each other for comparison. The $^{\tilde{A}}0$ DLIF is also displayed. Its 0-0 peak is shifted to overlap with $\tilde{B}$ state, in order to demonstrate a case without influence of the nearby states. In both SrOPh and SrOPh-d$_5$, many features from the $^{\tilde{A}}12^1 33 ^1$ state can be identified in the $^{\tilde{B}}0$ DLIF spectra, meaning some amount of collision-induced relaxation into this state must occur. However, in SrOPh-$d_5$,unique features from the $^{\tilde{A}}21^1 33 ^1$ state can be identified, indicating a different mechanism by vibronic interactions.}
    \label{fig:LIF}
\end{figure}

Figure \ref{fig:LIF}a shows the experimentally measured 2D spectra at energies near the $^{\tilde{B}}0$ excitation.
Fluorescence occurs predominantly around two energies corresponding to vertical relaxations between the same vibrational levels of the excited state $\tilde{A}/\tilde{B}$ and the ground state $\tilde{X}$. 
As explained in the Methods section, this feature arises from diagonal transitions between the same vibrational levels of the electronic states involved.
As such, we are able to determine the dominant electronic state components for each fluorescence peak.
For SrOPh, the $^{\tilde{B}}0$ peak is observed at the excitation wavelength ($\lambda_{air}$) 655.90 nm (15,249 cm$^{-1}$).
At a slightly lower energy of 655.15 nm (15266 cm$^{-1}$) excitation, an 
$\tilde{A}$-state peak is observed and assigned as $^{\tilde{A}}12^1 33 ^1$.
The vibrational mode of this level is assigned based on the computed vibrational frequency as well as our previous DLIF measurements~\cite{lao_laser_2022} ($^{\tilde{A}}12^1 33 ^1$ corresponds to $^{\tilde{A}}2^1 3 ^1$ in the old notation used in our previous work).
This combination mode appears because the transitions into the individual components $^{\tilde{A}}12^1$ and $^{\tilde{A}}33^1$ are favorable.
Note, however, this state does not have the correct symmetry to mix the $\tilde{A}$ and $\tilde{B}$ states and is absent in the KDC simulations.
We observe this state likely because of collisional relaxation from 
nearby $^{\tilde{B}}0$.

In SrOPh-d$_5$, the vertical excitation energy into the $^{\tilde{B}}0$ state is located at 656.00 nm (15,246 cm$^{-1}$), differing only slightly compared to SrOPh.
But the deuteration shifts all vibrational frequencies toward lower values, and the $^{\tilde{A}}12^1 33 ^1$ excitation is now located at 656.90 nm (15,225 cm$^{-1}$).
More importantly, we observe an additional excitation peak at 655.90 nm (15,249 cm$^{-1}$) near the $^{\tilde{B}}0$ energy.
This additional peak is of particular interest and is assigned to the vibronically coupled $^{\tilde{A}}21^1 33^1$ mode by comparing with the theoretical results.
As calculations suggest, deuteration brings the $^{\tilde{A}}21^1 33^1$ level closer to $^{\tilde{B}}0$ , greatly enhancing the coupling between the two vibronic states.
As such, excitation into $^{\tilde{A}}21^1 33^1$ borrows intensity from the neighboring strong $^{\tilde{B}}0$ excitations.
Given the large 0-0 transition probability, the transition into $^{\tilde{A}}21^1 33^1$ can be directly observed.
This mixing is much weaker in SrOPh because of a larger energy gap between the two vibronically coupled states, resulting in much weaker transition that cannot be resolved.

\subsubsection{Assignment of the vibrational modes}

The assignment of the vibrational modes of these excitation peaks are further confirmed by inspecting their DLIF spectra. 
However, since these peaks are close in energy, they may overlap with neighboring peaks. 
Therefore, we carefully compare their DLIF spectra by plotting all the nearby excitations together, as shown in Figure \ref{fig:LIF}b. 
For SrOPh, many features of the lower-energy $^{\tilde{A}}12^1 33^1$ indeed show up in the $^{\tilde{B}}0$ DLIF spectrum.
These features include the vertical relaxation from $^{\tilde{A}}12^1 33^1$ into the same vibration of the ground state, as well as the vibration-changing decays with losing or gaining one vibration quantum in the $\nu_{12}$ mode. 
When these features are excluded, the $^{\tilde{B}}0$ spectra match perfectly with those of the $^{\tilde{A}}0$ state.
We suspect that the strong intensity of the $^{\tilde{A}}12^1 33^1$ peaks results from the small energy gap with $^{\tilde{B}}0$, which greatly enhances the probability of collisional relaxations into this state.
In our previous work\cite{lao_laser_2022}, these $^{\tilde{A}}12^1 33^1$ features are mistakenly assigned to the $^{\tilde{A}}0$ state due to their similarities to the vertical relaxation wavelength. 
However, a slight difference between the $^{\tilde{A}}12^1 33^1 \rightarrow \text{}^{\tilde{X}}12^1 33^1 $ and $^{\tilde{A}}0 \rightarrow \text{}^{\tilde{X}}0$ transition frequencies can be observed from the 2D DLIF (see Fig. S1 in the SI).
Therefore, these peaks must be assigned to $^{\tilde{A}}12^1 33^1$. 
We observe similar $^{\tilde{A}}12^1 33^1$ features in the SrOPh-d$_5$ $^{\tilde{B}}0$ DLIF spectrum as well. 

For the SrOPh-d$_5$ DLIF spectra, a more important feature is the $^{\tilde{A}}21^1 33^1$ excitation, where several additional DLIF peaks only appear at this excitation energy. 
Despite the same set of collisional-induced $^{\tilde{A}}12^1 33^1$ peaks, three additional peaks, located at 63, 306, and 540 $\text{cm}^{-1}$, are absent in the nearby $^{\tilde{B}}0$ excitation. 
Therefore, we attribute these peaks solely to the $^{\tilde{A}}21^1 33^1$ state. 
Among the three peaks, the 306 $\text{ cm}^{-1}$ one has the largest intensity; it corresponds to the vertical relaxation to the same vibration of the ground state. 
The other two peaks are much weaker; they correspond to gaining or losing one vibrational quantum of the $\nu_{21}$ mode. 
This behavior exactly mirrors the nearby $^{\tilde{A}}12^1 33^1$ peaks, where the $\Delta \nu_{12} = 0,\pm1$ transitions are observed.
A complete assignment of the peaks in the DLIF spectra for SrOPh and SrOPh-d$_5$ can be found in the SI.

\begin{figure}[h!]
    \centering
    \includegraphics[width=0.8\linewidth]{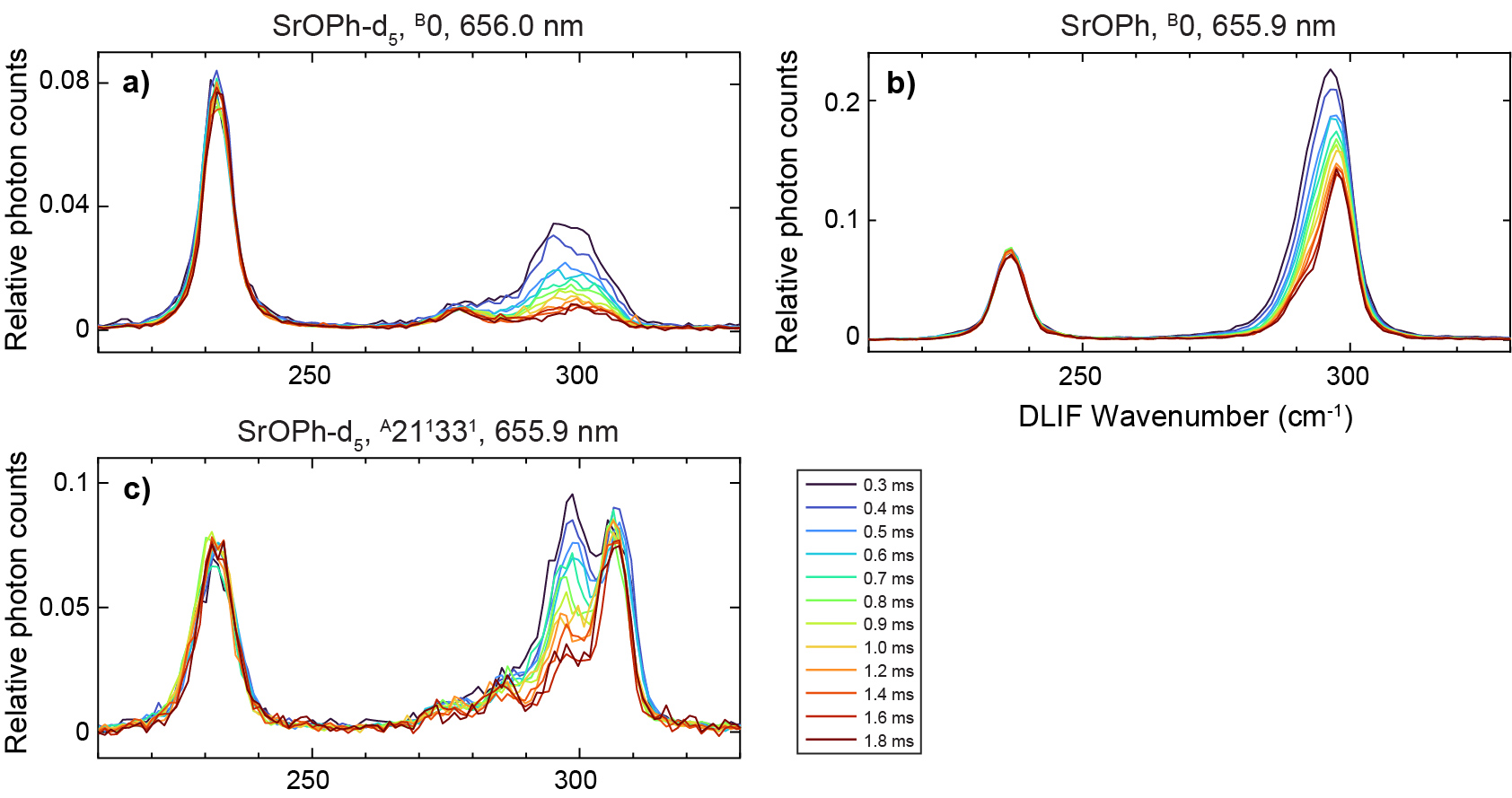}
    \caption{The dependence of the peaks with respect to the ablation-detection delays from 0.3 to 1.8 ms. By varying the experimental conditions, we are able to identify the peaks with changing relative intensities, meaning they do not originate from the same state we excite to. The character of these time-varying peaks agrees with our assignment in Figure \ref{fig:LIF}. To aid  comparison, the spectra are normalized to the  $^{\tilde{B}}0 \rightarrow \text{}^{\tilde{X}}0$  peak, which is not shown.}
    \label{fig:delay}
\end{figure}

The $\tilde{A}$ state relaxation peaks can be further confirmed by repeating DLIF measurements with varying experimental conditions. 
If the $^{\tilde{A}}12^1 33^1$ features are truly induced by collisional relaxations from the $^{\tilde{B}}0$ state, their intensities should respond differently to external changes.
This is observed when we vary the delay time between the laser ablation and the excitation. 
Typically, this delay is set to be $>1.0$ ms to allow enough time for hot molecules generated from laser ablation to cool down by colliding with the buffer gas. 
At earlier delay times, more hotter molecules are present; therefore, more collisions occur.\cite{skoff_diffusion_2011, hutzler_buffer_2012}
Indeed, the $^{\tilde{A}}12^1 33^1$ peaks have a much higher relative intensity at shorter delays.
Figure \ref{fig:delay} shows  the delay-time variation of the DLIF peak intensities 
relative to the $^{\tilde{B}}0 \rightarrow \text{}^{\tilde{X}}0$  transition for SrOPh and SrOPh-d$_5$. 
For both molecules, the strong variation of the $^{\tilde{A}}12^1 33^1$  peak at $\sim295 \text{ cm}^{-1}$ confirms that they are not from the same original state as $^{\tilde{B}}0$.
However, for the SrOPh-d$_5$ $^{\tilde{A}}21^1 33^1$ peak at $\sim306 \text{ cm}^{-1}$, this delay variation is basically the same as for $\ket{\tilde{B} ,\nu_0}$, which suggests that the population of this state is induced by state mixing rather than collisions. 
We observe similar temporal variation  for the vibrational-changing peaks of $^{\tilde{A}}12^1 33^1$ as well, which confirms our peak assignments in Figure \ref{fig:LIF}. 
However, because of the weaker intensity of the $^{\tilde{A}}21^1 33^1$ excitation, the intensity differences in its vibrational-changing peaks are too small to 
observe reliably.

\subsubsection{Deduction of the effective coupling strength}

Based on the relative intensities of the $^{\tilde{A}}21^1 33^1$ and $^{\tilde{B}}0$ relaxation peaks, a rough estimate of an effective coupling strength can be deduced using the intensity borrowing model\cite{lefebvre2004spectra,zhu_extending_2024}. A complete description of these couplings is more complex---the simulations based on the KDC Hamiltonian reveal that this coupling is induced by a second-order effect and is affected by many parameters. In a simplified model, considering vibronic coupling between only two states, the vibronically mixed eigenstates of the Hamiltonian are given as
\begin{eqnarray}
\ket{+} = c_{A+}\ket{\tilde{A},\nu_{21}\nu_{33}} + c_{B+}\ket{\tilde{B},\nu_{0}},
\protect\label{eq:plus}\\
\ket{-} = c_{A-}\ket{\tilde{A},\nu_{21}\nu_{33}} + c_{B-}\ket{\tilde{B},\nu_{0}},
\protect\label{eq:minus}
\end{eqnarray}
where $\ket{+}$ and $\ket{-}$ denote the two coupled states and $c_{A/B,+/-}$ denote the corresponding coefficients. In the experiments, the relative intensities between the two vertical relaxations $^{\tilde{A}}12^1 33^1 \rightarrow \text{}^{\tilde{X}}12^1 33^1 $ and $^{\tilde{B}}0 \rightarrow \text{}^{\tilde{X}}0 $ can be measured reliably. If we assume similar Franck--Condon overlap from vibration-preserving transitions, then the relative ratios of the peak intensities depend on the excitation process, which is determined by the coefficient of the $^{\tilde{B}}0$ component. As such,
\begin{equation}
\frac{I_+}{I_-} = \left|\frac{c_{B+}}{c_{B-}}\right|^2=\frac{1+\Delta E^0 / \Delta E}{1-\Delta E^0 / \Delta E},
\end{equation}
where $\Delta E =\sqrt{{\Delta E^0}^2+4H_{12}^2}$ is the energy gap between the perturbed states, given by the coupling strength $H_{12}$ and the uncoupled energy gap $\Delta E^0$. 
In SrOPh-d$_5$, using the separation of the energy state $\Delta E = 3 \text{ cm}^{-1}$ and an intensity ratio of $\sim0.04$, we obtain an effective coupling strength $H_{12} \simeq0.5 \text{cm}^{-1}$. 
If we use this coupling strength for SrOPh together with the computed $\Delta E^0 = 20 \text{ cm}^{-1}$, the calculated relative intensity is $<0.001$, which explains the absence of $^{\tilde{A}}12^1 33^1$ peak in SrOPh. 
Upon a careful scan of the 2D DLIF of SrOPh, it seems that a tiny $\tilde{A}$ state peak can be identified at 655.45 nm (15,259 cm$^{-1}$) by comparing its relative intensity with the $\tilde{B}$ state peak (see Fig.~\ref{fig:SI_SrOPh} in the SI for more details). If we take the same coupling strength, this suggests $\Delta E \simeq 10 \text{ cm}^{-1}$ and an intensity ratio of $\sim0.01$ that is observable. However, we note that these intensities are extremely weak and comparison between relative ratios of weak features should be taken with caution.

The experimentally determined effective coupling strength $H_{12}\simeq0.5 \text{cm}^{-1}$ between $\tilde{A}$ and $\tilde{B}$ is a second-order effect arising from the diabatic couplings between  the $\tilde{A}-\tilde{C}$ and $\tilde{B}-\tilde{C}$ states. This effective value has no equivalent in the theoretical model and should not be compared directly. Nonetheless, this second-order effect appears to be much weaker than the first-order couplings between the $\tilde{A}-\tilde{C}$ and $\tilde{B}-\tilde{C}$ states ($\sim150 \text{ cm}^{-1}$). However, with appropriate vibronic states lying close in energy, it is still possible to introduce state-mixing and create additional decay channels. 

\subsubsection{Comparison with theory}

\begin{figure}[h!]
    \centering
    \includegraphics[width=\linewidth]{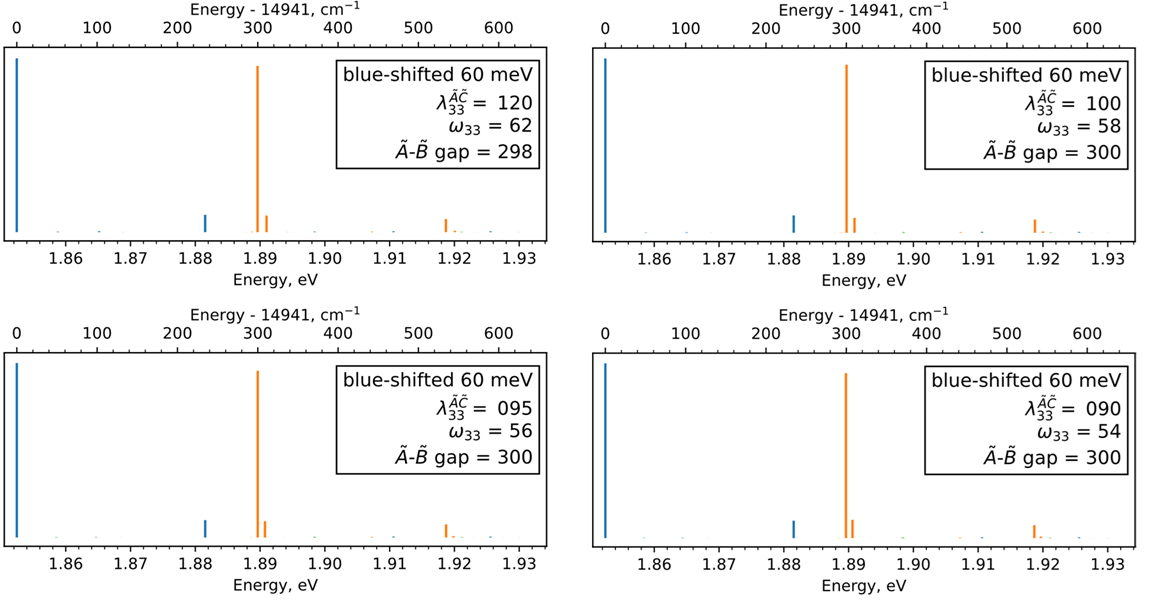}
    \caption{SrOPh-d$_5$. Spectra simulated by varying theoretical parameters. Two of the parameters, the frequency of the lowest mode $\omega_{33}$ and the $\tilde{A}-\tilde{C}$ coupling strength of this mode $\lambda_{33}$ are varied.}
    \label{fig:Theo_vary}
\end{figure}

Compared to the simulations, experimentally observed $^{\tilde{A}}12^1 33^1$ peaks are less intense.
The separation of the two vibronically coupled peaks are also smaller in the experiments. 
In order to improve the agreement between the theory and experiment, we carried out several simulations by varying the two most relevant theoretical parameters, the frequency of the lowest mode $\omega_{33}$ and the strength of
the $\tilde{A}-\tilde{C}$ coupling of this mode, $\lambda_{33}$. Figure~\ref{fig:Theo_vary} shows the
results.
A satisfactory match can be obtained by reducing the coupling strength down to $\sim$60\% and shifting the vibrational frequency to compensate.
These simulations demonstrate that the KDC Hamiltonian is capable of explaining the complex experimental spectrum
as resulting from the vibronic interactions. Quantitative agreement can be achieved by manual adjustments to the model parameters. The need for manual adjustment stems from the approximations introduced in order to achieve the numerical tractability of the problem.
This study demonstrates the importance of vibronic coupling in determining the optical cycling and laser cooling properties. 
Future theoretical developments on modeling such complex systems with more vibrational modes to resolve even finer details are important.

\section{Conclusions}
\label{sec:conculsions}

In summary, we have characterized vibronic interactions between the first and second electronically excited states,
$\tilde{A}$ and $\tilde{B}$, of the OCC-functionalized SrOPh molecule.
The coupling results in state mixing between the closely separated vibronic levels, enhancing the intensities of the normally weak or forbidden transitions. 
By comparing SrOPh and its deuterated analogue with different vibrational frequencies, we were able to unravel
the details of these couplings by varying the relative energy separations between the two coupled states. 
A coupling strength on the order of $\sim0.5$~cm$^{-1}$ is measured for this specific case.
The coupling we observed is a second-order effect mediated through the linear coupling of the higher $\tilde{C}$ electronic state. This mechanism is similar to the spin--orbit vibrational coupling term in the linear SrOH molecule~\cite{zhang_accurate_2021, zhang_intensity-borrowing_2023}. 
Whereas such coupling is two orders of magnitude weaker than the direct (linear) vibronic coupling mechanism, with appropriate energy levels nearby, they contribute to extra decay channels that cannot be ignored for optical cycling and laser cooling purposes.

Although our current work only investigates the SrOPh molecule, given the closely lying vibronic levels in many similar molecules (for example, Ca/SrOPh-X derivatives), we have also observed abnormally large intensities of $\tilde{A}$ state peaks in their DLIF spectra~\cite{zhu_functionalizing_2022}.
We anticipate that similar vibronic  effects can contribute and play an important role.
However, because of the difficulties in obtaining the deuterated species for many complex molecules,
we did not take measurements with more candidates. 
CaOPh is less prone to such coupling mechanism because of the smaller separation between its $\tilde{A}$ and $\tilde{B}$ states ($\sim130 \text{ cm}^{-1}$).
The energy gap is only able to support the first two vibrational (Ca-O-C bending) modes, greatly limiting the vibronic states available for coupling. 
However, linear coupling between the $\tilde{A}$ and $\tilde{C}$ states may still lead to additional undesirable branching pathways.
It is clear that, when considering optical cycling and laser cooling of complex molecules, theoretical efforts must go beyond the Born--Oppenheimer approximation to accurately predict the decay channels.

\section*{Supplementary Material}
The supplementary materials includes the full 2D spectra of SrOPh and SrOPh-d$_5$, the assignments of the fluorescence peaks, and the details of the computational protocol.

\section*{Acknowledgments}
We thank Professor David Nesbitt from JILA for helpful conversations. AIK and PW are grateful to the late Professor John F. Stanton for his guidance in mastering KDC simulations. 
This work was supported by the NSF Center for Chemical Innovation Phase I (Grant No. CHE-2221453) and NSF PHY-2110421. WCC and ERH acknowledge institutional support from NSF OMA-2016245. 

\bibliography{Cstate,pawels_library,allrefs}

\pagebreak
\widetext
\begin{center}
\textbf{\large Supplementary Information for:}
\linebreak
\textbf{\large Unraveling vibronic interactions in molecules functionlized with optical cycling centers}
\end{center}
\setcounter{equation}{0}
\setcounter{figure}{0}
\setcounter{table}{0}
\setcounter{page}{1}
\setcounter{section}{0}
\makeatletter
\renewcommand{\theequation}{S\arabic{equation}}
\renewcommand{\thefigure}{S\arabic{figure}}
\renewcommand{\thetable}{S\arabic{table}}

\section{Assignment of the peaks in the DLIF spectra}

\begin{table}[h!]
\caption{SrOPh: Assignment of the DLIF spectra}
\begin{tabular}{ll}
\hline
Energy ($\text{cm}^{-1}$) & Transition     \\ \hline
54            & $^{\tilde{B}}0\rightarrow \text{}^{\tilde{X}}33^1$ \\
60            & $^{\tilde{A}}12^1 33^1\rightarrow \text{}^{\tilde{X}}33^1$  \\
236           & $^{\tilde{B}}0\rightarrow\text{}^{\tilde{X}}12^1$  \\
291           & $^{\tilde{B}}0\rightarrow\text{}^{\tilde{X}}12^1 33^1$   \\
297           & $^{\tilde{A}}12^1 33^1\rightarrow \text{}^{\tilde{X}}12^1 33^1$ \\
474           & $^{\tilde{B}}0\rightarrow\text{}^{\tilde{X}}11^1$ \\  
531           & $^{\tilde{A}}12^1 33^1\rightarrow \text{}^{\tilde{X}}12^2 33^1$  \\  
599           & $^{\tilde{B}}0\rightarrow\text{}^{\tilde{X}}10^1$  \\  \hline
\end{tabular}
\end{table}

\begin{table}[h!]
\centering
\caption{SrOPh-d$_5$: Assignment of the DLIF spectra}
\begin{tabular}{ll}
\hline
Energy ($\text{cm}^{-1}$) & Transition             \\ \hline
48            & $^{\tilde{A}}12^1 33^1\rightarrow \text{}^{\tilde{X}}33^1$ \\
52            & $^{\tilde{B}}0\rightarrow \text{}^{\tilde{X}}22^1$  \\
63            & $^{\tilde{A}}21^1 33^1\rightarrow \text{}^{\tilde{X}}33^1$  \\
231           & $^{\tilde{B}}0\rightarrow \text{}^{\tilde{X}}12^1$  \\
276           & $^{\tilde{B}}0\rightarrow \text{}^{\tilde{X}}12^1 33^1$   \\
296           & $^{\tilde{A}}12^1 33^1\rightarrow \text{}^{\tilde{X}}12^1 33^1$   \\
304           & $^{\tilde{A}}21^1 33^1\rightarrow \text{}^{\tilde{X}}21^1 33^1$    \\
357           & $^{\tilde{A}}12^1 33^1\rightarrow \text{}^{\tilde{X}}12^1 33^2$ \\
463           & $^{\tilde{B}}0\rightarrow\text{}^{\tilde{X}}11^1$ \\  
534           & $^{\tilde{A}}12^1 33^1\rightarrow \text{}^{\tilde{X}}12^2 33^1$  \\ 
540           & $^{\tilde{A}}21^1 33^1\rightarrow \text{}^{\tilde{X}}21^2 33^1$  \\ 
591           & $^{\tilde{B}}0\rightarrow\text{}^{\tilde{X}}10^1$  \\  \hline
\end{tabular}
\end{table}
\pagebreak

\section{Detailed 2D LIF spectroscopy of $\text{SrOPh}$ and $\text{SrOPh-d}_5$}

\begin{figure}[h!]
    \centering
    \includegraphics[width=\linewidth]{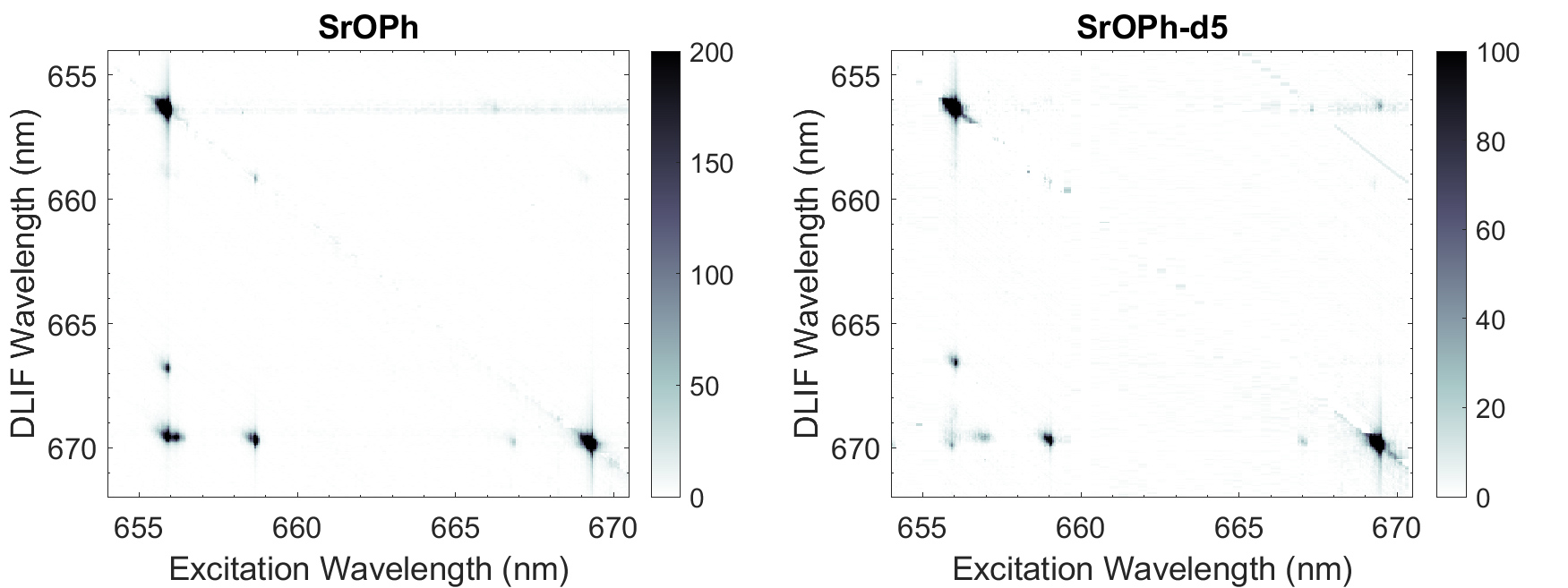}
    \caption{Complete 2D LIF spectra of SrOPh and $\text{SrOPh-d}_5$. A small difference in the fluorescence wavelength of different vibrational excitations of the $\tilde{A}$ state can be observed.}
    \label{fig:SI_2D_full}
\end{figure}
\pagebreak

\section{Determination of the weak $^{\tilde{A}}21^1 33^1$ peak in $\text{SrOPh}$}

\begin{figure}[h!]
    \centering
    \includegraphics[width=\linewidth]{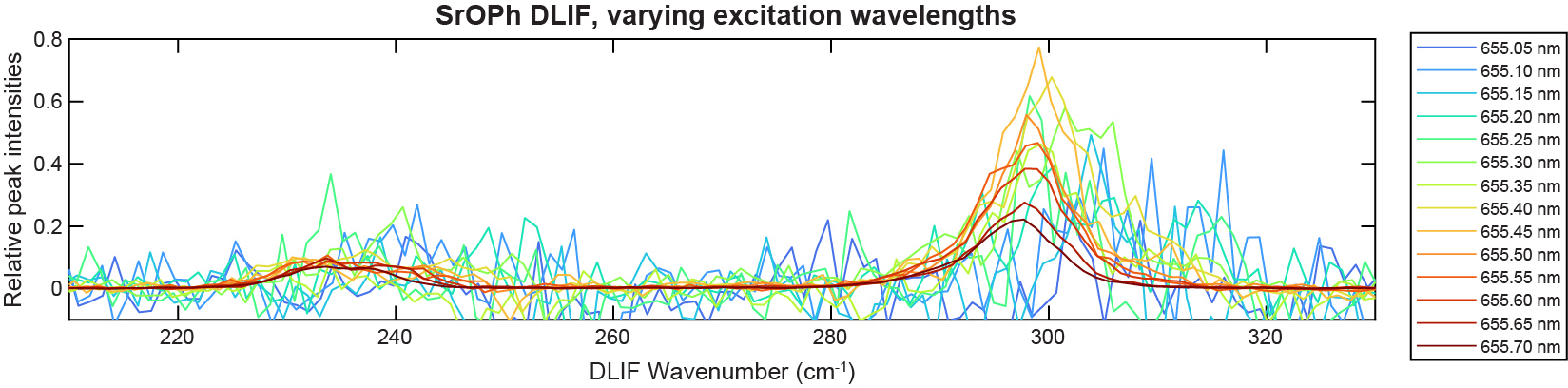}
    \caption{SrOPh. DLIF spectra for varying excitation wavelengths. Only the portion near the $^{\tilde{B}}0 \rightarrow \text{}^{\tilde{X}}12^1$ relaxation peak and $\tilde{A} \rightarrow \tilde{X}$ vertical relaxation is shown. The peak intensities are normalized with respect to the $\tilde{B} \rightarrow \tilde{X}$ 0-0 intensity at each wavelength. At 655.45 nm, an extra peak can be identified on the right side of the $\tilde{A} \rightarrow \tilde{X}$ vertical relaxation peaks. We believe this corresponds to the $^{\tilde{A}}21^1 33^1$ state, which is weakly enhanced by coupling with the $^{\tilde{B}}0$ state. However, because of its weak intensity, a similar analysis by varying delay time is not performed}
    \label{fig:SI_SrOPh}
\end{figure}

The $^{\tilde{A}}21^1 33^1$ state is observed to be $\sim3 \text{ cm}^{-1}$ above the $^{\tilde{B}}0$ state in SrOPh-d$_5$. In SrOPh, we expect a larger energy separation between the two states and, therefore, a much weaker state mixing. Whereas it is hard to directly observe such a weak peak with a much more intense $^{\tilde{B}}0$ state nearby, we provide some side evidence of the presence of $^{\tilde{A}}21^1 33^1$  peak in SrOPh by comparing the relatively intensities. Figure~\ref{fig:SI_SrOPh} shows the DLIF spectra at 655.05-665.70 nm by normalizing between the $\tilde{A}$ state and $\tilde{B}$ state peaks. As can be seen, an intensity of the relative intensity of the $\tilde{A}$ state peak is observed at 655.45 nm. This peak is also slightly red-shifted, which agrees with the behavior of the $^{\tilde{A}}21^1 33^1$ peak in SrOPh-d$_5$. As such, we believe this peak corresponds to the $^{\tilde{A}}21^1 33^1$ state excitation of SrOPh.

\clearpage
\section{Details of the computational protocol}
\label{si:sec:comp_details}

\begin{table}[h!]
    \caption{SrOPh. 
        Vertical excitation energies in cm$^{-1}$. The first row
        presents the energy of the $\tilde{A}$ state while the rows for the
        $\tilde{B}$ and $\tilde{C}$ states show the energy gap above the
        $\tilde{A}$ state. Calculated using EOM-EA-CCSD and correlating all
        electrons, with pseudopotential for the core Sr electrons ECP28MDF.
    }
    \label{tab:sroph_vertical_cbs}
  \center
    \begin{tabular}{|c|c|c|c|c|}
        \hline
    $E ^{(\alpha)}$ & aug-cc-pwCVDZ & aug-cc-pwCVTZ & aug-cc-pwCVQZ & CBS \\
        \hline
    $\tilde{A}$ &   14288 &   14501 &   14512 &   14520 \\
    $\tilde{B}$ &    +195 &    +169 &    +167 &    +165 \\
    $\tilde{C}$ &   +1966 &   +1668 &   +1615 &   +1577 \\
        \hline
    \end{tabular}
\end{table}

\begin{table}[h!]
    \caption{SrOPh. 
        Vertical excitation energies in cm$^{-1}$. Corrections
        due to the triples correlation ($\Delta$EOM-EA-CCSD$^*$) and spin--orbit
        couplings.
    }
    \label{tab:sroph_vertical_dT_SOC}
  \center
    \begin{tabular}{|c|c|c|c|c|}
        \hline
    $E ^{(\alpha)}$ & EOM-CCSD/CBS & $\Delta$EOM-CCSD$^*$ & SOC & final \\
        \hline             
    $\tilde{A}$ &  14520       &  +40 & -73 & 14487 \\
    $\tilde{B}$ &  14685       &  +24 & +53 & 14762 \\
    $\tilde{C}$ &  16097       & -209 & +10 & 15898 \\
        \hline
    \end{tabular}
\end{table}

The ground-state geometry was optimized using EOM-EA-CCSD/cc-pVDZ[H,C,O]/cc-pwCVDZ-DK2[Sr]. The calculations of the harmonic frequencies, potential expansion coefficients
$\kappa$ and linear diabatic couplings $\lambda$, were carried out at the same
level of theory.

The calculation of the vertical excitation energies $E^{(\alpha)}$ were
carried out using a composite scheme: the EOM-EA-CCSD energies were calculated
using a series of basis sets and extrapolated to the complete basis set (CBS)
limit. The effect of higher excitations was added by calculating the difference
between the EOM-EA-CCSD$^*$ and EOM-EA-CCSD energies, and the difference was
applied to the energies as a triples correction. The spin--orbit couplings (SOC)
were calculated between the $\tilde{X}$, $\tilde{A}$, $\tilde{B}$, and
$\tilde{C}$ states, the SOC Hamiltonian was constructed and diagonalized (see the
SI of Ref.~\cite{Khvorost:dualOCC:2024} for details) to yield the final
values of the vertical excitation energies.

Table~\ref{tab:sroph_vertical_cbs} 
lists the vertical excitation energies and the CBS limit. The main
observation is that the quadruple-$\zeta$-quality basis is within
$10$~cm$^{-1}$ from the CBS limit for the $\tilde{A}$ and $\tilde{B}$ state and
about five times as far for the $\tilde{C}$ state.
Table~\ref{tab:sroph_vertical_dT_SOC} lists the corrections due to higher
excitations
and the SOCs. The magnitude of the triples correction is of the order of
tens of wavenumbers for the $\tilde{A}$ and $\tilde{B}$ states, and, again, 
much larger for the $\tilde{C}$ state. The SOCs repel the $\tilde{A}$ and
$\tilde{B}$ state by more than $100$~cm$^{-1}$. The SOCs do not shift the
energy of the $\tilde{C}$ state by much, however, the $\tilde{C}$ state may
interact stronger with the higher states, which are not included in this
model. The right-most column of Table~\ref{tab:sroph_vertical_dT_SOC} shows the
final values of this composite scheme, these values are also plotted in
Fig.~\ref{fig:SrCouplings} of the main manuscript.

\end{document}